\documentclass[12pt]{article}

\usepackage{caption}
\usepackage{amsmath,amsfonts,amsthm,amssymb}
\usepackage{subfigure}
\usepackage{setspace}
\usepackage{Tabbing}
\usepackage{lastpage}
\usepackage{extramarks}
\usepackage{chngpage}
\usepackage[usenames,dvipsnames]{color}
\usepackage{graphicx,float,wrapfig}

\newcommand{\lp}{\left(}\newcommand{\rp}{\right)}

\def\IZ{\rlx\hbox{\sf Z\kern-.4em Z}}
\def\IR{\rlx\hbox{\rm I\kern-.18em R}}
\def\IC{\rlx\hbox{\,$\inbar\kern-.3em{\rm C}$}}
\def\one{\hbox{{1}\kern-.25em\hbox{l}}}

\begin{document}
\numberwithin{equation}{section}

\begin{titlepage}

\begin{center}
{\bf Collective coordinate approximation to the scattering of solitons\\ in the (1+1) dimensional NLS model}
\end{center}

\vspace{.5cm}

\begin{center}
{H. E. Baron~$^{\dagger}$, G. Luchini~$^{\star}$\\ and\\ W. J. Zakrzewski~$^{\dagger}$}

\vspace{.2 in}
\small

\par \vskip .2in \noindent
$^{(\star)}$ Departamento de Ci\^encias Naturais,\\
 Universidade Federal do Esp\'irito Santo,\\
CEP 29932-540, S\~ao Mateus-ES, Brazil,\\
gabriel.luchini@ceunes.ufes.br

\par \vskip .2in \noindent
$^{(\dagger)}$~Department of Mathematical Sciences,\\
  Durham University, Durham DH1 3LE, U.K.\\
h.e.baron@durham.ac.uk,\\
w.j.zakrzewski@durham.ac.uk

\end{center}

\par \vskip 1 in

\begin{abstract}
We present a collective coordinate approximation to model the dynamics of two interacting nonlinear Schr\"odinger (NLS) solitons. We discuss the accuracy of this approximation by comparing our results with those of the full numerical simulations and find that the approximation is remarkably accurate when the solitons are some distance apart, and quite reasonable also during their interaction.
\end{abstract}
\end{titlepage}

\section{Introduction}
The nonlinear Schr\"odinger equation (NLS) is an important model in mathematical physics, with applications in many fields which includes nonlinear optics, plasma physics, biophysics and Bose-Einstein condensates (BEC's). Interactions between NLS solitons is particularly important; for example in soliton-based optical communications the NLS equation describes information transfer in optical fibres \cite{Hase2000}, and soliton interactions fundamentally limit the capacity of these communication systems \cite{Kunz1999}.

As the NLS equation is integrable its exact soliton solutions can be found analytically via the inverse scattering transform \cite{Zakh1971} (see {\it e.g.} \cite{Akt}). However, given the rather involved nature of this approach and the complicated form of these solutions and the fact that they hold only for the exact form of the NLS equation it is useful to look at other approaches
to this problem. This is particularly true if one wants to get a `physical feeling' about the forces governing the scattering
of solitons {\it i.e.} to see whether they are attractive or repulsive and how they depend on the various parameters of the solutions and how they respond to small perturbations of these solutions or the equation itself.

Hence, the equation has also been studied numerically \cite{Koda1987}, \cite{Roth1992}, \cite{Gord1983}, \cite{Mits1987} and an attempt has been made to introduce a collective coordinate approximation to a two soliton field configuration \cite{Zou1994}. Several other papers have also looked at NLS solitons perturbed 
by external fields or in interaction with them \cite{extra} but though very interesting, these papers have not approximated the dynamics of the system 
of solitons by a full Lagrangian based collective coordinate model \cite{Mant1982}, which has recently been shown \cite{book}, \cite{Paul} (in relativistic models) to be a very good 
approximation for the investigation of soliton dynamics.

Having performed some numerical 
simulations of the scattering of two solitons in a class of modified NLS models \cite{Zakr2012} we have started thinking of a collective coordinate approximation to this process and we have found the paper by Zou and Yan \cite{Zou1994}.
As this paper does not present many explicit results we have modified its approach a little and have looked at the interaction of two solitons in some detail. We have found that the collective approach, which is expected to describe the properties of the solitons when they are far apart from each other, works quite well even when the solitons are close together and so may be a somewhat unexpectedly good  approximation to the description of the two soliton scattering at all times. Thus our paper discusses
this approximation and its validity for a class of models based on the NLS in (1+1) dimensions.

This paper is organised as follows: in section 2 we give a brief introduction to the NLS model, its basic symmetries and its 1-soliton solution. In section 3, for completeness,  we say a few words about the collective coordinate approximation in general, and in section 4 we present our 2-soliton approximation ansatz (based on \cite{Zou1994}) and use it to determine the equations of motion for our collective coordinates. We have solved these equations numerically  using the 4th order Runge Kutta method,  and in section 5, we present some of our results. Some further comments and  conclusions are given in section 6.

\section{The model}
The non-relativistic Lagrangian describing the dynamics of the  NLS field $\psi(t,x)$ and its complex conjugate $\psi^\ast(t,x)$ is given by
\begin{equation}
\label{lag}
\mathcal{L}= \int dx\; \frac{i}{2}\left( \psi^\ast\partial_t \psi - \psi \partial_t \psi^\ast\right) - \partial_x\psi^\ast \partial_x\psi + \eta \vert \psi \vert^4 .
\end{equation}
Variation of this Lagrangian with respect to $\psi^\ast(t,x)$ gives us

\begin{equation}
\label{eom}
i\partial_t \psi = -\partial_x^2\psi - 2 \eta \vert \psi \vert^2\psi,
\end{equation}
which is the NLS equation for $\psi(t,x)$ (variation of the Lagrangian with respect to $\psi(t,x)$ gives the complex conjugate of \eqref{eom} which is the NLS equation for $\psi^\ast(t,x)$).

Solutions to \eqref{eom} with boundary conditions $|\psi|_{x=-\infty}=|\psi|_{x=\infty}$; $\partial_x \psi\rightarrow 0$ as $x\rightarrow \pm \infty$ have conserved Noether charges as a result of the symmetries of the action. 

Thus the invariance of the action under time translations gives the energy conservation:
\begin{equation}
\label{energy}
E= \int_{-\infty}^{\infty} dx \; \lp |\partial_x \psi|^2-\eta |\psi|^4 \rp.
\end{equation} 

Conservation of momentum results from the invariance of the action under space translations:
\begin{equation}
\label{momentum}
P = i\int_{-\infty}^{+\infty} dx \; \lp \psi^\ast \partial_x \psi-\psi \partial_x \psi^\ast \rp.
\end{equation}
And, finally,  the internal $U(1)$ symmetry of the action, $\psi \rightarrow e^{i\alpha} \psi$ for a constant $\alpha$, gives the conservation of the normalisation
\begin{equation}
\label{norm}
N= \int_{-\infty}^{+\infty} dx \; | \psi|^2.
\end{equation}

As is well known for $\eta=1$, \eqref{eom} has the 1-soliton solution (called `bright soliton')
\begin{equation}
\label{sol_bright}
\psi = \frac{b}{\cosh{\left[b\left( x-vt-x_0 \right)\right]}}e^{i\left[ \left(b^2 - \frac{v^2}{4}  \right)t+\frac{v}{2}x +\delta\right]},
\end{equation}
where $b$, $v$ and $x_0$ are real parameters of the solution. This solution is clearly defined up to an overall constant phase due to the $U(1)$ symmetry of \eqref{lag}. It describes a soliton moving with velocity $v$, which at $t=0$ is positioned 
at $x_0$. The parameter $b$, which describes the `width' of the soliton, is related to $N$ and so is, in fact, fixed.

\section{The collective coordinate approximation}

For integrable systems exact solutions can be found via the inverse scattering transform (IST); however IST is confined to integrable models so for non integrable systems, or when one wants to study perturbations of integrable models, other methods must be used to find approximate solutions or to understand what is really going on. In such cases, one can perform numerical
simulations ({\it i.e.} solve the equations numerically but this is often very time consuming) or use other approximate methods. One of such methods is the collective coordinate approximation \cite{Sanc1998}. This approximation reduces the infinite-dimensional problem to a coupled set of ODEs for the collective coordinates by focusing on the motion of the solitons themselves,
and so retaining only the variables which describe the solitons. Of course, this approximation neglects all radiative corrections and so is valid only if these corrections are small; this is true when the solitons are far apart from each other. 
When the solitons begin to interact with each other the approximation becomes less accurate (as some radiation is sent out and 
the solitons are mutually distorted). However, it may happen that these distortions are well described by the well chosen
collective coordinates and that the radiation effects are small. This is, in fact, what we have found in our work as will
be described in the next few sections.

The general idea of the collective coordinate approximation is to start with a static solution $\psi(x,q_1,...,q_n)$. Of course, if $\psi(x,q_1,...,q_n)$ is a static solution then the total energy of the solution does not depend on the values
of the parameters ({\it i.e.} for all values of these parameters the energy is the same). Some of these parameters describe physical properties of the solitons, like their position {\it etc}. If we change the field configuration describing the solitons
the energy will be larger so that in the field space we have  low energy valleys in the directions of the parameters of the solutions with the slopes described 
by the other modifications of the fields. 

 Consider now moving solitons. For small velocities of the solitons tangential to the field space their motion would be easiest along the valleys descibed by the parameters of the static solutions
as other changes (going up the slopes) would require larger increases of the energy. 
Hence, for small velocities it makes sense to approximate the dynamics of the solitons by the parameters of the static solution becoming functions of $t$; {\it i.e.} $q_i=q_i(t)$, and assuming that these parameters contain all the solitons' dynamics.

These assumptions are reasonable for the relativistic field theories and the collective coordinate approximation (also called the moduli space approximation) has been studied 
in detail in many papers (see {\it e.g.} \cite{book} or \cite{Paul} for the study of the Sine Gordon case). The approximation is very good and reproduces the results of the full simulations of such systems very well indeed.

The NLS model is a little different as its equations of motion involve first derivative
with respect to time and the energy \eqref{energy} does not contain a kinetic
contribution. Moreover, the model has stationary and not static solutions (see \eqref{sol_bright}). Clearly $x_0+vt$ denotes the position of the soliton
moving with velocity $v$ at time $t$, so a natural collective coordinate would be $\xi(t)=x_0+vt$. However, the soliton possesses also a moving phase, which has to be taken into account in any collective coordinate approximation. As the $x$ dependent 
 part is proportional to the velocity of the moving soliton it makes sense 
to introduce a collective coordinate $\mu(t)$ which initially takes the value of $\frac{v}{2}$; there is also the overall constant phase $\delta$ in \eqref{sol_bright}.

How does one obtain the equations for the collective coordinates? This is discussed 
in great detail in \cite{book} where it is shown that one takes relevant collective 
coordinate approximation ansatz and puts it into the expression for the action. One then integrates
out all relevant spatial degrees of freedom (in our case $x$) and obtains a Lagrangian
for the collective coordinates $q_i(t)$. In our case, as the full Lagrangian 
is given by \eqref{lag},  the resultant Lagrangian will involve $q_i(t)$ and will be linear in $\dot q_i(t)$, and from it we can determine the first order equations 
for $q_i(t)$. Of course, these equations are much easier to solve than the original 
equation  \eqref{eom} and it is often easier to understand the dynamics.
At the same time, however, the collective coordinate model is only an approximation 
which does not capture some aspects of the dynamics {\it e.g.} any radiation effects that often accompany scattering processes. Moreover, there is often an issue of
which collective coordinates to use and whether they are sufficient to capture
the main features of the dynamics.

\section{The 2-soliton configuration}

Here we construct a set of collective coordinates for the study of the scattering of two solitons. In the NLS case there exists an explicit expression 
for the two moving solitons. However, this expression is not very transparent and 
when the solitons are far apart it reduces to the superposition approximation ansatz which we will make below.
Moreover, when we go beyond the pure NLS model ({\it i.e.} modify it slightly)
we do not have explicit expressions and we are obliged to start by constructing a sensible approximation 
ansatz. So, our work also involves a check for the suitability of our approximation ansatz.

The motivations for our approximation ansatz is the observation that when the solitons are far away from each other each one of them is well described
by \eqref{sol_bright}. The overlap between them is very small so we take the two soliton field in the form of a superposition of two
independent solitons {\it i.e.} we take

\begin{equation}
\label{ansatz}
\psi=\psi_1+\psi_2.
\end{equation}
 Where $\psi_1$ and $\psi_2$ are solutions of (\ref{eom}) when they are far apart. Following from Zou and Yan, \cite{Zou1994}, we assume that the two solitons are of equal height, constant width, and move symmetrically around their centre of mass. 
So we take $\psi_1=\varphi_1 e^{-i\theta_1}$ and $\psi_2=\varphi_2e^{i\theta_2}$ where
$$
\varphi_1=\frac{a(t)}{\cosh{\left(b(x+\xi(t))\right)}}, \qquad \theta_1=\mu(t) \left( x+ \xi(t)\right) - b^2t-\lambda(t) - \delta_1,
$$
$$
\varphi_2=\frac{a(t)}{\cosh{\left(b(x-\xi(t))\right)}}, \qquad \theta_2=\mu(t) \left( x- \xi(t)\right) + b^2t+\lambda(t) + \delta_2,
$$
and then treat $a(t), \ \xi(t), \ \mu(t)$ and $\lambda(t)$ as our collective coordinates.

\begin{figure}[!ht]
\centering
\includegraphics[scale=1.1]{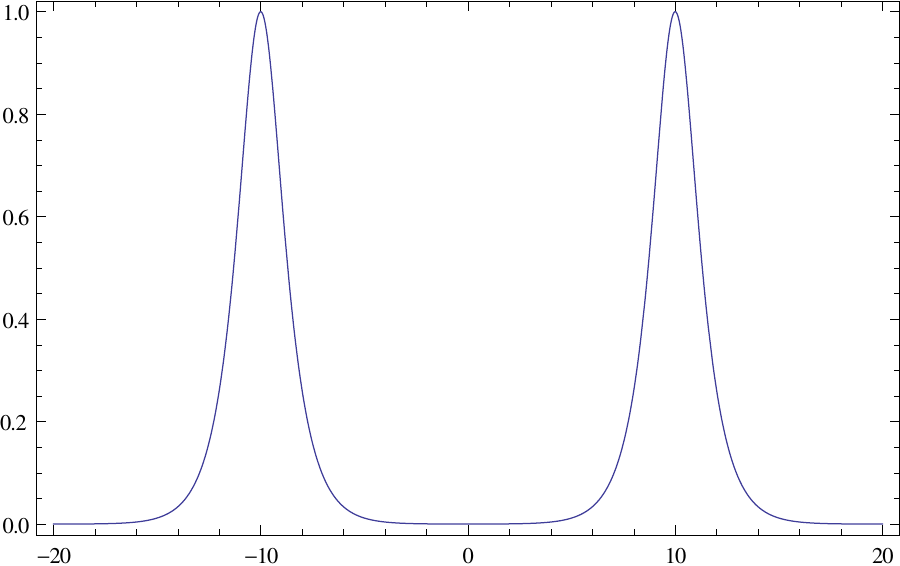}
\caption{Plot of $\psi=\psi_1+\psi_2$ against $x$, for $\psi$ the 2-soliton approximation of the NLS model.}
\label{fig:2}
\end{figure}

 This approximation ansatz models two lumps with relative phase $\delta=\delta_2-\delta_1$ and relative distance $2\xi$ and so corresponds to two 1-soliton solutions when $|\xi| \rightarrow \infty$. In figure \ref{fig:2} we present a plot of $\psi=\psi_1+\psi_2$ at $t=0$ with $\xi=10$, $\mu=0.1$, $b=1$, $\lambda=0$ and $\delta_1=\delta_2=0$.

\subsection{Effective Langrangian for our collective coordinates}

To construct the effective Lagrangian for our collective coordinates we put our approximation ansatz \eqref{ansatz} into our Lagrangian 
\eqref{lag}, this yields an effective Lagrangian density which can be written in terms of the non-interacting part $\mathcal{L}_0$ and the interacting part $\mathcal{L}_{12}$. 

Introducing $\omega_1\equiv x+\xi$ and $\omega_2 \equiv x-\xi$, the non-interacting part becomes
\begin{eqnarray*}
\mathcal{L}_0&=&a^2\left( \mu \dot{\xi} - b^2-\dot{\lambda}-\mu^2 \right) \left(\frac{1}{\cosh^2(b\omega_1)}+\frac{1}{\cosh^2(b\omega_2)}\right)-a^2b^2\left(\frac{\tanh^2(b\omega_1)}{\cosh^2(b\omega_1)}+\frac{\tanh^2(b\omega_2)}{\cosh^2(b\omega_2)}\right)\\
&+& a^4 \left(\frac{1}{\cosh^4(b\omega_1)}+\frac{1}{\cosh^4(b\omega_2)}\right)+a^2\dot{\mu}\left(\frac{\omega_1}{\cosh^2(b\omega_1)}-\frac{\omega_2}{\cosh^2(b\omega_2)}\right),
\end{eqnarray*}
where dot denotes the differential with respect to time. Integrating this over all space gives us the effective Lagrangian of free solitons
\begin{equation}
L_0=\frac{4a^2\mu\dot{\xi}}{b}-\frac{16a^2b}{3}-\frac{4a^2\mu^2}{b}-\frac{4a^2\dot{\lambda}}{b}+\frac{8a^4}{3b}.
\end{equation}

Defining $\theta_1+\theta_2=2\mu x + \delta_2- \delta_1 \equiv 2\mu x + \delta \equiv \Delta$, the interacting Lagrangian density becomes
\begin{eqnarray*}
\mathcal{L}_{12}&=&-a^2b\left(\dot{\xi}+2\mu\right)\left(\frac{\sinh(b\omega_1)}{\cosh(b\omega_2)\cosh^2(b\omega_1)}+\frac{\sinh(b\omega_2)}{\cosh(b\omega_1)\cosh^2(b\omega_2)}\right)\sin\Delta\\
&+&2a^2\left(\mu^2+\dot{\mu}\xi +\mu \dot{\xi}-b^2- \dot{\lambda}\right)\frac{\cos\Delta}{\cosh(b\omega_1)\cosh(b\omega_2)}\\
&-&2a^2b^2\frac{\sinh(b\omega_1)\sinh(b\omega_2)}{\cosh^2(b\omega_1)\cosh^2(b\omega_2)}\cos\Delta\\
&+&4a^4\left(\frac{1}{\cosh^3(b\omega_1)\cosh(b\omega_2)}+\frac{1}{\cosh^3(b\omega_2)\cosh(b\omega_1)}\right)\cos\Delta\\
&+&\frac{2a^4}{\cosh^2(b\omega_1)\cosh^2(b\omega_2)}\cos\left(2\Delta\right)+\frac{4a^4}{\cosh^2(b\omega_1)\cosh^2(b\omega_2)},
\end{eqnarray*}
which, when integrated over space, and after some rearranging yields
\begin{eqnarray*}
L_{12}&=&\left(\dot{\mu}\xi-\mu^2 - \dot{\lambda}+ \frac{4 a^2\mu^2}{b^2}\right)\frac{4 \pi a^2 \sin(2\mu\xi)\cos\delta}{b \sinh(\frac{\pi\mu}{b})\sinh(2b\xi)}+\left(1-\frac{2 a^2}{b^2}\right)\frac{8 \pi a^2 b\sin(2\mu\xi)\cos\delta}{\sinh(\frac{\pi\mu}{b})\sinh^3(2b\xi)}\\
&+&
\left(\frac{2 a^2}{b^2}-1\right)\frac{8\pi\mu a^2\cos(2\mu\xi)\cosh(2b\xi)\cos\delta}{\sinh(\frac{\pi \mu}{b})\sinh^2(2b\xi)}+32a^4\xi \frac{\cosh(2b\xi)}{\sinh^3(2b\xi)}-\frac{16a^4}{b\sinh^2(2b\xi)}\\
&+&\frac{8\pi a^4\cosh(2b\xi)\sin(4\mu\xi)\cos(2\delta)}{b\sinh(\frac{2\pi \mu}{b})\sinh^3(2b\xi)}-\frac{16 \pi a^4 \mu\cos(4\mu\xi)\cos(2\delta)}{b^2\sinh(\frac{2\pi\mu}{b})\sinh^2(2b\xi)}.
\end{eqnarray*}

The integrals given here have been evaluated using the residue theorem; some of these calculations are presented in detail in the Appendix.

\subsection{Equations of motion}

Next we determine the equations for our collective coordinates. 
First we note that the total Lagrangian is given by

\begin{eqnarray*}
L&=&\frac{4 a^2}{b}\left( \mu \dot{\xi}-\frac{4 b^2}{3}-\mu^2-\dot{\lambda}+\frac{2 a^2}{3}\right)+\left(\dot{\mu}\xi-\mu^2 - \dot{\lambda}+ \frac{4 a^2\mu^2}{b^2}\right)\frac{4 \pi a^2 \sin(2\mu\xi)\cos\delta}{b \sinh(\frac{\pi\mu}{b})\sinh(2b\xi)}\\
&+&\left(1-\frac{2 a^2}{b^2}\right)\frac{8 \pi a^2 b\sin(2\mu\xi)\cos\delta}{\sinh(\frac{\pi\mu}{b})\sinh^3(2b\xi)}+
\left(\frac{2 a^2}{b^2}-1\right)\frac{8\pi\mu a^2\cos(2\mu\xi)\cosh(2b\xi)\cos\delta}{\sinh(\frac{\pi \mu}{b})\sinh^2(2b\xi)}\\
&+&32a^4\xi \frac{\cosh(2b\xi)}{\sinh^3(2b\xi)}-\frac{16a^4}{b\sinh^2(2b\xi)}+\frac{8\pi a^4\cosh(2b\xi)\sin(4\mu\xi)\cos(2\delta)}{b\sinh(\frac{2\pi \mu}{b})\sinh^3(2b\xi)}\\
&-&\frac{16 \pi a^4 \mu\cos(4\mu\xi)\cos(2\delta)}{b^2\sinh(\frac{2\pi\mu}{b})\sinh^2(2b\xi)}.
\end{eqnarray*}
This expression agrees with the Lagrangian given in Zou and Yan's paper \cite{Zou1994} if we take their approximation by neglecting higher order terms of $\mu$, $\lambda$ and their $t$ derivatives. 

From our full Lagrangian we can calculate the Euler-Lagrange equations for our collective coordinates $a(t), \ \xi(t), \ \mu(t)$ and $\lambda(t)$.

For $\lambda$ we have
$$
\frac{d}{dt}\frac{\partial L}{\partial \dot{\lambda}}-\frac{\partial L}{\partial \lambda}=0 \rightarrow \frac{d}{dt} \left(\frac{4a^2}{b}\left(1+\frac{\pi \sin(2\mu\xi)\cos\delta}{\sinh(\frac{\pi\mu}{b})\sinh(2b\xi)}\right)\right)=0,
$$
which implies that 
$$
\frac{4a^2}{b}\left(1+\frac{\pi \sin(2\mu\xi)\cos\delta}{\sinh(\frac{\pi\mu}{b})\sinh(2b\xi)}\right)=\text{constant},
$$
is a conserved quantity corresponding to the normalisation $N$. So we can write
$$
N=\int_{-\infty}^{+\infty}dx\;\vert \psi \vert^2=\frac{4a^2}{b}\left(1+\frac{\pi \sin(2\mu\xi)\cos\delta}{\sinh(\frac{\pi\mu}{b})\sinh(2b\xi)}\right)\equiv N_0 + N_{12},
$$
where $N$ has been split into interacting and non-interacting parts.

Next we fix $N$, which is conserved and so does not depend on $t$, by putting solitons initially far apart, {\it i.e.} taking $x_0$ very large. In our 2-soliton approximation $\psi_1$ and $\psi_2$ are 1-soliton solutions for the solitons far apart, if we compare this to the 1-soliton solution \eqref{sol_bright} we see that for our solitons initially far apart $\mu \approx -\frac{v}{2}$, $\xi\approx x_0 - vt$ and $a\approx b$, and therefore $N_{12}\approx 0$, $N_{0}\approx 4b$.

Then we have
\begin{equation}
\label{eq_lamb}
a^2=\frac{b^2}{1+\frac{\pi \sin(2\mu\xi)\cos\delta}{\sinh(\frac{\pi\mu}{b})\sinh(2b\xi)}}\equiv \frac{b^2}{1+\omega},
\end{equation}
where we have defined $\omega\equiv\frac{\pi \sin(2\mu\xi)\cos\delta}{\sinh(\frac{\pi\mu}{b})\sinh(2b\xi)} $ for convenience.

Equation (4.3) can be used to eliminate $a(t)$ from the equations of motion for $\mu(t)$ and $\xi(t)$, giving a system of coupled first order equations involving $\mu$, $\xi$, their derivatives and $\dot{\lambda}$. The dependence in $\dot{\lambda}$ can be eliminated if we use the equation of motion for $a(t)$, leaving us with
\begin{eqnarray*}
F_1(\mu,\xi)\dot{\mu}+G_1(\mu,\xi)\dot{\xi}+H_1(\mu,\xi)=0,\\
F_2(\mu,\xi)\dot{\mu}+G_2(\mu,\xi)\dot{\xi}+H_2(\mu,\xi)=0.
\end{eqnarray*}
Finally we solve these to derive the system of equations
\begin{equation}
\label{syst}
\dot{\mu}=\frac{G_1H_2-G_2H_1}{F_1G_2-F_2G_1},\qquad \dot{\xi}=\frac{F_2H_1-F_1H_2}{F_1G_2-F_2G_1}.
\end{equation}

We write the right hand side of the expression for $\dot{\mu}$ as $R(\mu,\xi)$, and differentiate the expression with time to get
$\ddot{\mu}=\dot{R}$. Multiplying this by $\dot{\mu}$ and integrating over time gives a conserved quantity $E$
\begin{equation}
\label{cons}
\frac{\dot{\mu}^2}{2}=\frac{R^2}{2}+E,
\end{equation}
where $E$ is determined by the initial conditions. Similarly we can do this for the expression for $\dot{\xi}$ to get
\begin{equation}
\label{cons_1}
\frac{\dot{\xi}^2}{2}=\frac{P^2}{2}+\text{\~{E}},
\end{equation}
so we have two energy-like conservation formulas. If we consider $\frac{{\dot{\mu}}^2}{2}$ to be like kinetic energy, $-\frac{R^2}{2}$ to be like a potential and $E$ to be like total energy then we can plot potential curves as $-R^2$ up to a constant (we take this constant to be the square of the initial velocity), see figure \ref{fig:potentials}.

\section{Results}
In our work we have used the fourth-order Runge-Kutta method to solve numerically our system of equations (\ref{syst}). Each 1-soliton configuration, $\psi_1$ and $\psi_2$, possesses a $U(1)$ symmetry so we can choose each phase arbitrarily and consider the dependence on their phase difference $\delta$. In our analysis we have considered only small values of velocity ($\dot \xi$) describing the initial motion of the solitons towards each other, as the collective coordinate approximation is a good approximation for slowly moving solitons.

\begin{figure}[!ht]
\centering
\includegraphics[trim = 2cm 2cm 11cm 11cm, scale=0.7]{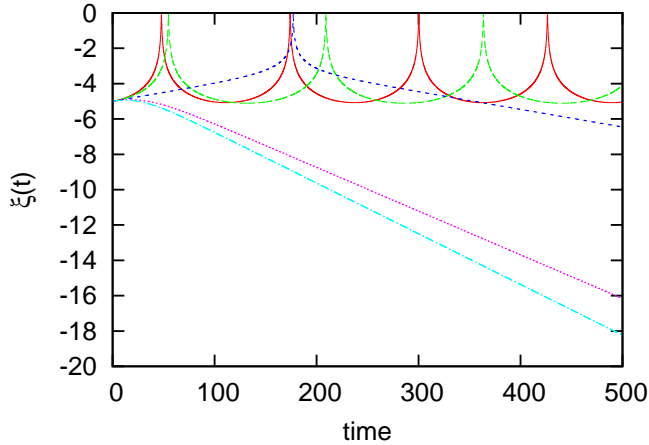}
\caption{The relative position of the solitons for different values of $\delta$, the phase difference between the two solitons: $\delta=0$ (red line), $\delta=\frac{\pi}{4}$ (green line), $\delta=\frac{\pi}{2}$ (dark blue line), $\delta=\frac{3 \pi}{4}$ (pink line) and $\delta=\pi$ (light blue line).  }
\label{fig:Trajectories}
\end{figure}
Our simulations of the collective coordinate approximation have shown that the interaction between the solitons depends on their initial phase difference and their velocity at the time of interaction. Solitons with the same initial phase ($\delta=0$) attract each other the most and, if their velocity is sufficiently small, they become trapped and oscillate around each other with constant frequency. Solitons with the opposite initial phase ($\delta=\pi$) are in the repulsive channel and so they repel each other. The attractive/repulsive forces vary continuously between $\delta=0, \pi$ with complex interactions taking place around $\delta=\frac{\pi}{2}$ where the solitons experience an initial attraction and so come together, then repel and move away from each other with a constant velocity. The range of interactions can be seen in figure \ref{fig:Trajectories} where the relative position between the solitons is plotted as a function of time, for a simulation with the initial distance $\xi=-5$, initial velocity $v=-0.01$ so that they are sent towards each other, and for $\delta=0$,  $\frac{\pi}{4}$, $\frac{\pi}{2}$,  $\frac{3\pi}{4}$ and $\pi$.

\begin{figure}[!ht]
\centering
\includegraphics[trim = 2cm 2cm 11cm 11cm, scale=0.7]{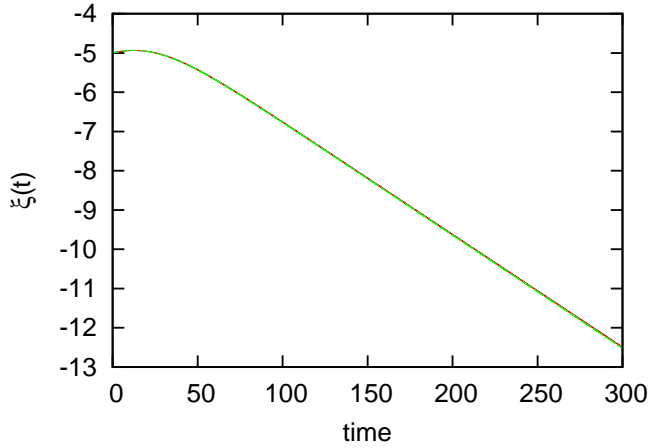}
\caption{The relative position of the solitons initially at $\xi=-5$, with an initial velocity $v=-0.01$ and phase difference $\delta=\pi$; results of the full simulation is the dashed line and the approximation is the solid line.}
\label{fig:k=2}
\end{figure}
 \begin{figure}[!ht]
\centering
\includegraphics[trim = 2cm 2cm 11cm 11cm, scale=0.7]{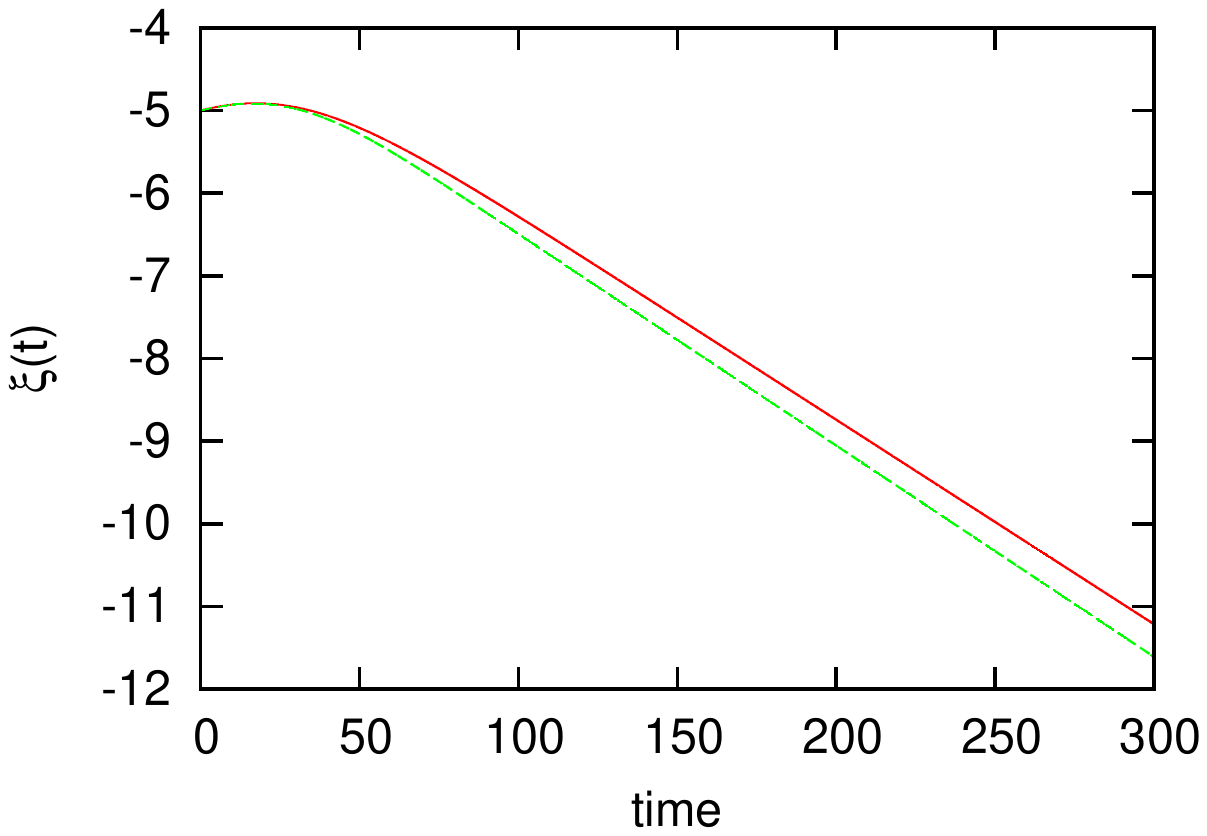}
\caption{The relative position of the solitons initially at $\xi=-5$, with an initial velocity $v=-0.01$ and phase difference $\delta=\frac{3\pi}{4}$; results of the full simulation is the dashed line and the approximation is the solid line.}
\label{fig:k=3/2}
\end{figure}
\begin{figure}[!ht]
\centering
\includegraphics[trim = 2cm 2cm 11cm 11cm, scale=0.7]{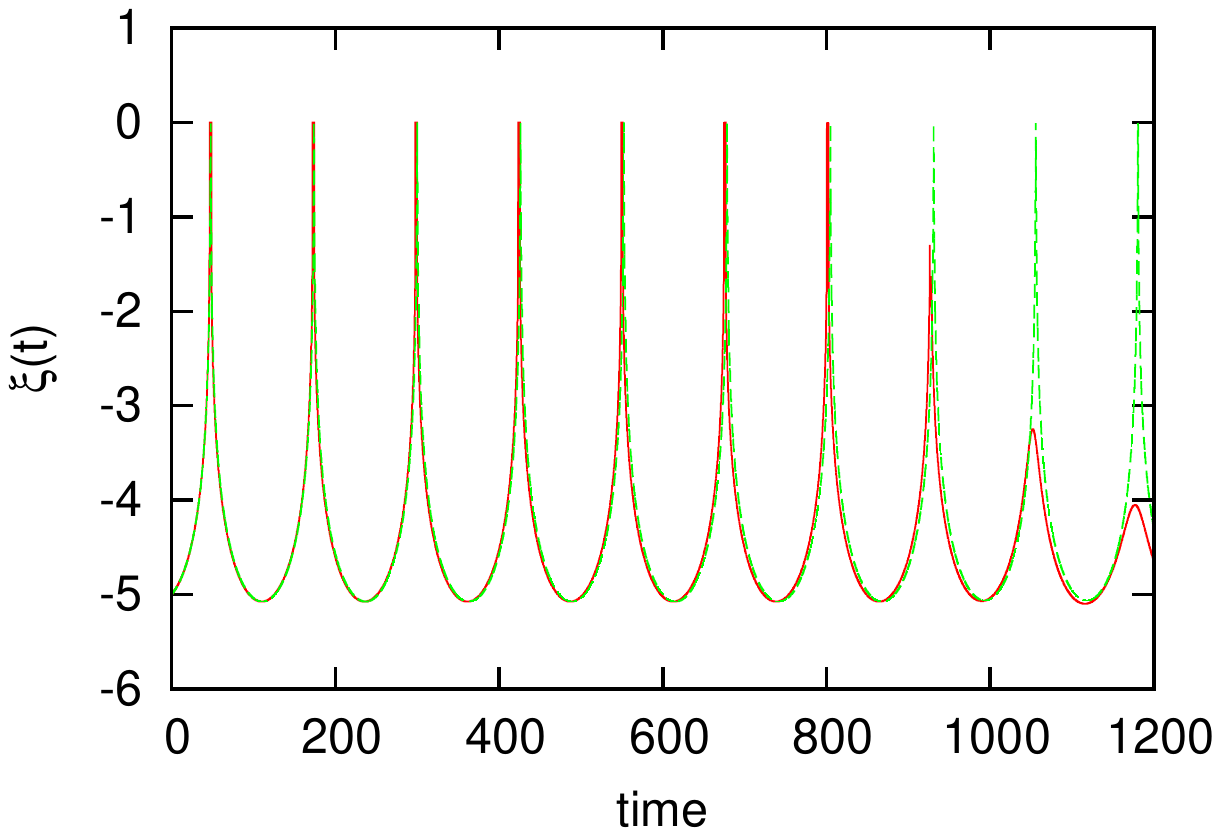}
\caption{The relative position of the solitons initially at $\xi=-5$, with an initial velocity $v=-0.01$ and phase difference $\delta=0$; results of the full simulation is the dashed line and the approximation is the solid line.}
\label{fig:k=0}
\end{figure}\begin{figure}[!ht]
\centering
\includegraphics[trim = 2cm 2cm 11cm 11cm, scale=0.7]{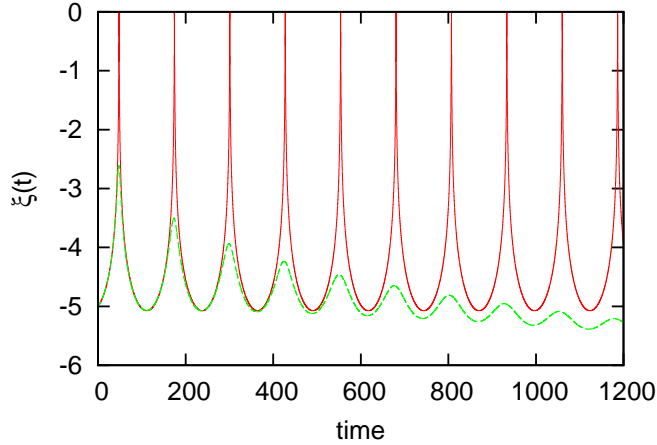}
\caption{The relative position of the solitons initially at $\xi=-5$, with an initial velocity $v=-0.01$ and phase difference $\delta=\frac{\pi}{32}$; results of the full simulation is the dashed line and the approximation is the solid line.}
\label{fig:k=1/16}
\end{figure}
Comparison of the approximation with the full simulation confirms the observed dependence of the soliton scattering on the initial phase difference between the solitons, $\delta$, and shows that the approximation describes the dynamics of the soliton scattering with varying levels of accuracy for different values of $\delta$. For $\delta=\pi$ the approximation is very accurate, this can be seen in figure \ref{fig:k=2} where the results of the full simulation and the approximation are both plotted for solitons initially at $\xi=-5$ and with an initial velocity $v=-0.01$ (so that they are sent towards each other), and with relative phase $\delta=\pi$. In the repulsive cases, $\delta\gtrsim \frac{\pi}{2}$, the results for the full simulation and the approximation are very close, see figure \ref{fig:k=3/2} where the full simulation and approximation results are compared for $\delta=\frac{3\pi}{4}$, and initial $\xi=-5$, $v=-0.01$ as before. However, for values of $\delta\lesssim\frac{\pi}{2}$ our collective coordinate approximation does not fully capture the soliton dynamics. For small values of $\delta$ in the full simulation the solitons initially attract and oscillate as in the approximation, but over time the oscillations weaken and the solitons start to repel each other. For $\delta=0$ the approximation remains accurate for a long time as the oscillations only start to decay at around $t=900$, see figure \ref{fig:k=0}. For small non zero values of $\delta$ the decay starts immediately and the approximation does not match the full simulation as well, though it does give a close approximation for the period of the oscillations, this can be seen in figure \ref{fig:k=1/16} where the results are compared for $\delta=\frac{\pi}{32}$. For values of $\delta$ closer to $\frac{\pi}{2}$ the attraction is so weak that the solitons only move towards each other for a short period of time before repelling away, this is different to the approximation where the solitons move together slowly and come on top of each other before slowly oscillating (or eventually repelling if initial velocity is too high). These differences could be due to the phase difference being a constant in our collective coordinate approximation but free to vary in time in the full simulation, therefore allowing solitons intially in an attractive channel to end up in a repulsive channel. This could be tested by changing the choice of collective coordinates to allow the solitons' phases to  vary separately in time.

\begin{figure}
  \centering
  \subfigure[]{\label{fig:potential1}\includegraphics[trim = 2cm 2cm 12cm 12cm, width=0.45\textwidth]{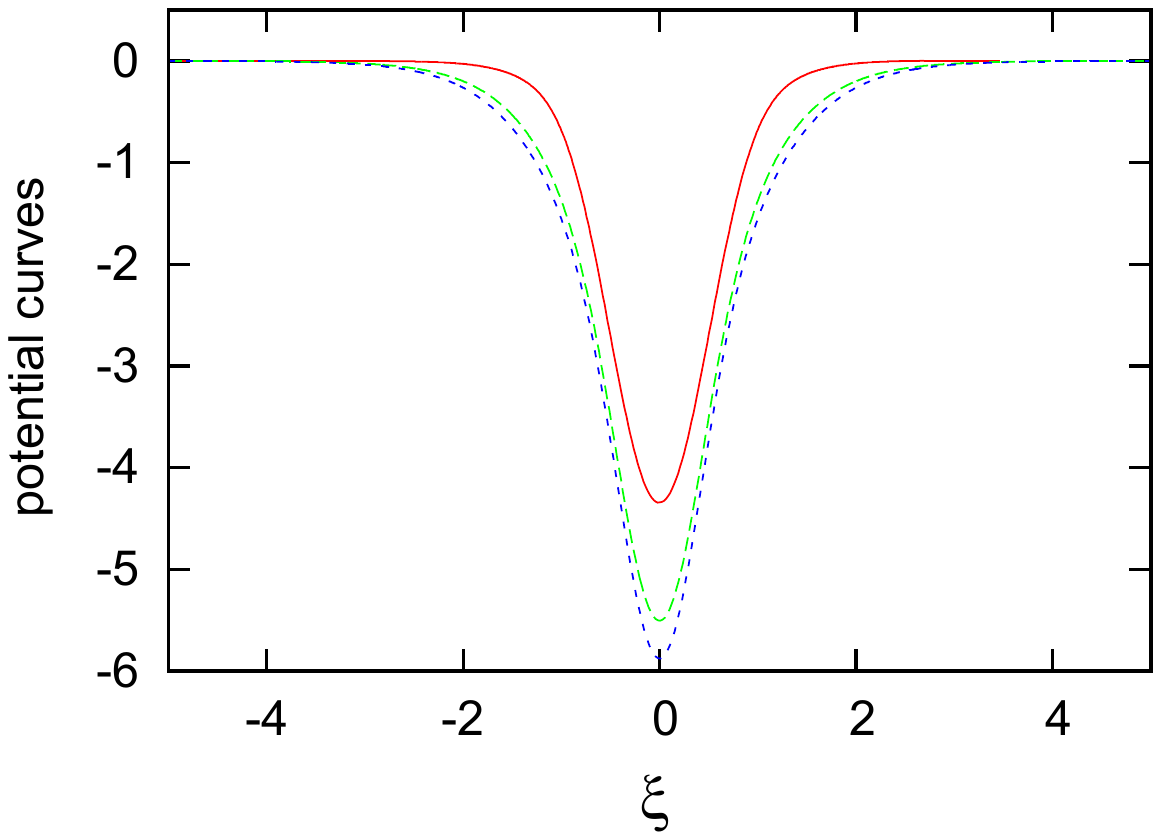}}                
  \subfigure[]{\label{fig:potential2}\includegraphics[trim = 2cm 2cm 12cm 11.2cm, clip, width=0.45\textwidth]{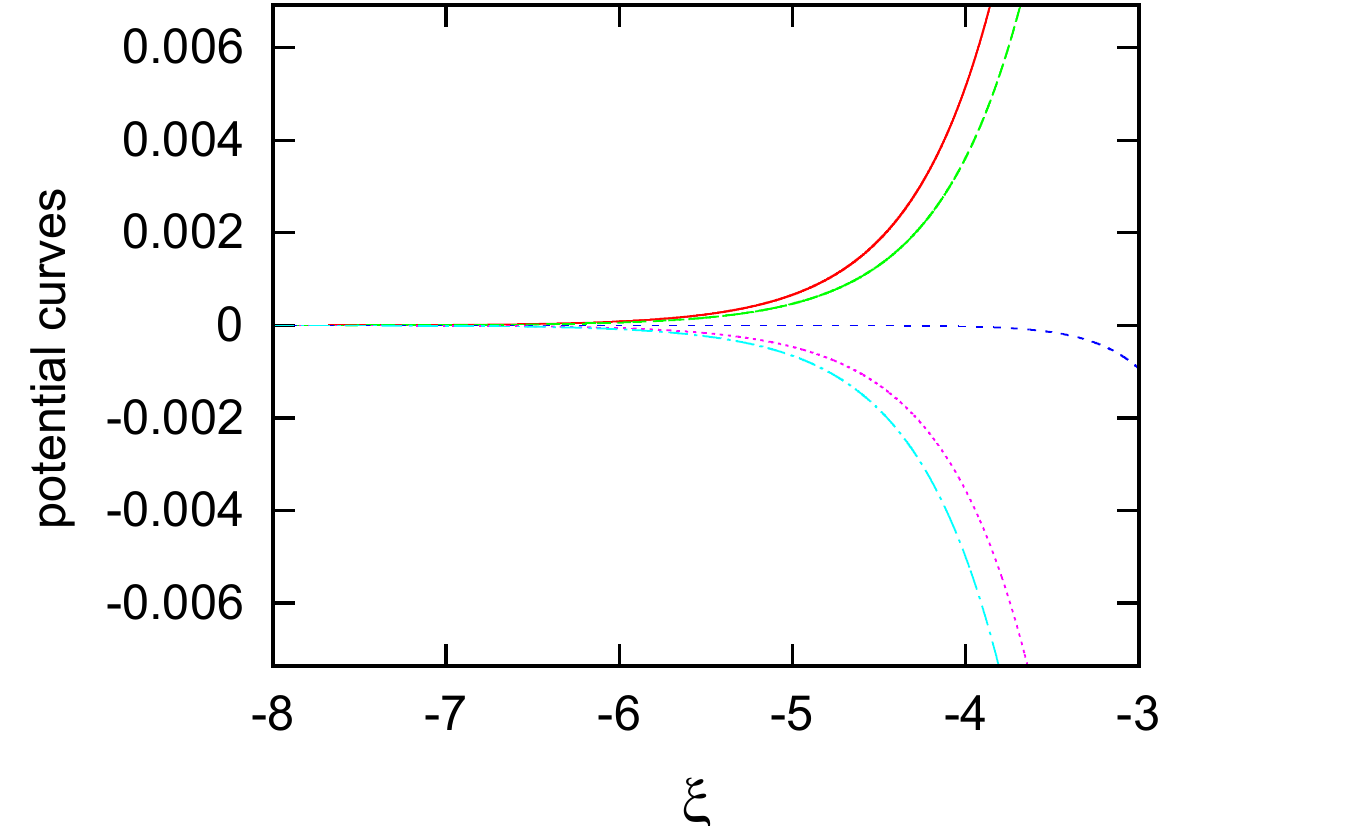}}
  \caption{Potential curves for solitons initially at $\xi=-10$ and $v=-0.1$ with a) from top to bottom $\delta=\frac{\pi}{2}$, $\frac{\pi}{4}$, $0$, and b) from top to bottom $\delta=\pi$, $\frac{3\pi}{4}$, $\frac{\pi}{2}$, $\frac{\pi}{4}$, $0$}
  \label{fig:potentials}
\end{figure}

\begin{figure}[!ht]
\centering
\includegraphics[trim = 2cm 2cm 11cm 11cm, scale=0.7]{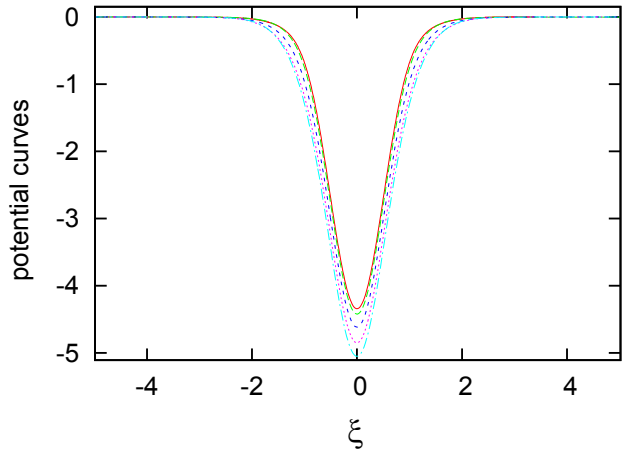}
\caption{Potential curves for solitons initially at $\xi=-10$, $\delta=\frac{\pi}{2}$ and from top to bottom $v=-0.000001$, $-0.5$, $-1$, $-1.5$, $-2$}
\label{fig:pot_v}
\end{figure}

We have confirmed our observations by considering the conserved quantity resulting from our expression for $\dot{\xi}$.  This we have done by interpreting (\ref{cons}) as an energy conservation formula so that we could consider the movement of solitons as the motion of a particle moving in a potential. In figure \ref{fig:potentials} we have plotted the potential curves for initial velocity $v=-0.1$, initial position $\xi=-10$, and various values of $\delta$. We see that $\delta=0, \pi$ do indeed correspond to the attractive and  repulsive potentials, respectively. Our potential curves are similar to those in Zou and Yan's results in \cite{Zou1994} but with a few differences as we have not made any approximations in our calculations. Firstly, our potential curves have a dependence on the initial velocity which is demonstrated in figure \ref{fig:pot_v} by plotting potential curves for $\delta=\frac{\pi}{2}$, initial position $\xi=-10$ and various values of initial velocity. Secondly our potential curves are more symmetric about $\delta=\frac{\pi}{2}$, {\it i.e.} in our results solitons with $\delta=\pi$ / $\delta=0$ feel repulsion/attraction at the same relative distance, whereas in Zou and Yan's results solitons with $\delta=\pi$ feel repulsion whilst further apart than solitons with $\delta=0$ feel attraction. Finally, our potential curve for $\delta=\frac{\pi}{2}$ is much more attractive than theirs for all values of the initial velocity (see figure \ref{fig:potential1}).

\section{Further comments and some conclusions}

In this paper we have presented a collective coordinate approximation (based on the modifcation of the approach of Zou and Yan \cite{Zou1994}) for the study of the dynamics of two interacting bright solitons in a NLS model and then we have used it to investigate these dynamics in some detail. We have observed that the initial relative phase between the solitons determines whether they feel an attractive or repulsive force towards each other, and for a small enough velocity the solitons can form a bound state and continue to oscillate around each other indefinitely. In comparing our results to those of full numerical simulations we had remarkable agreement in most cases, suggesting that our collective 
coordinate approximation can be used to reproduced the dynamics of the solitons even when the solitons are close together. We have also observed some discrepancies for small values of relative phase which we hope to be able to resolve in further work by adjusting our choice of collective coordinates. In addition to this we plan to continue our work by applying the method developed in this report to investigate various physically interesting perturbations to the NLS equation.
\section{Acknowledgements}
HB is supported by an STFC studentship. GL thanks Durham University for its hospitality and also thanks CNPq which provided financial support through the Science Without Borders program.

\newpage
\appendix
\section{Appendix: Calculation of integrals}
\begin{figure}[!ht]
\centering
\includegraphics[scale=0.5]{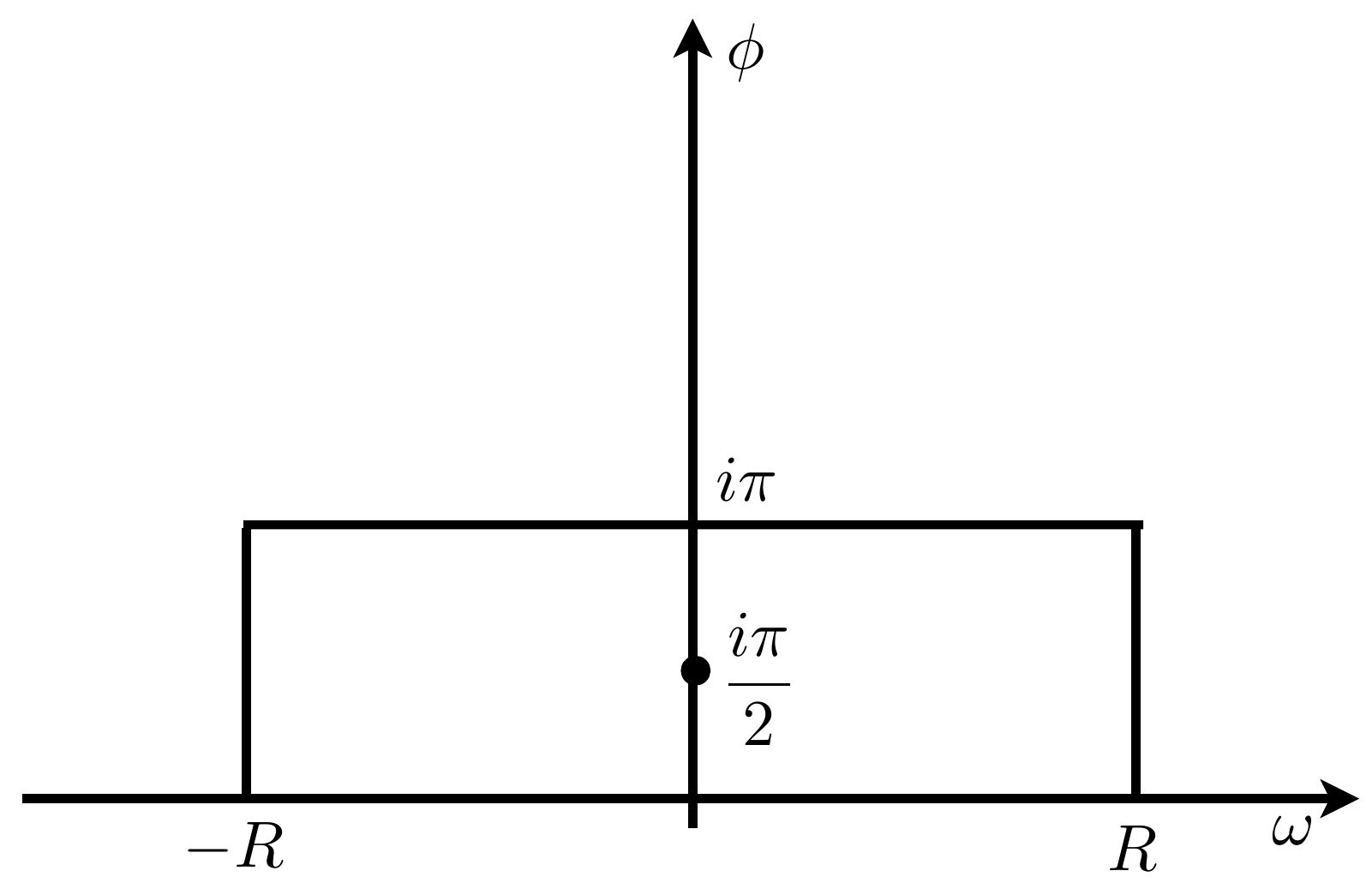}
\caption{Appropriate contour (called $C$) for all the integrals: only one of the infinitely many poles is picked.}
\label{fig:contour}
\end{figure}

Here we present a few details which show the way we have performed the calculations of the integrals in section 4.

\subsection{$I=\int_{-\infty}^{+\infty}\frac{dx}{\cosh^2(b(x+\xi(t)))\cosh^2(b(x-\xi(t)))}$}
Defining $\omega=b(x+\xi(t))$ we can write:
\begin{equation}
I=\int_{-\infty}^{+\infty}\frac{dx}{\cosh^2(b(x+\xi(t)))\cosh^2(b(x-\xi(t)))}=\frac{1}{b}\int_{-\infty}^{+\infty}\frac{d\omega}{\cosh^2(\omega)\cosh^2(\omega-2b\xi)}.
\end{equation}
Consider the following complex integral along the closed contour $C$ (see figure \ref{fig:contour}) in the plane $z=\omega+i\phi$
\begin{equation}
\oint_C f(z) dz=\oint_C \frac{z }{\cosh^2(z)\cosh^2(z-2b\xi)}dz.
\end{equation}
We have chosen our contour such that the integrand is analytic except for two second-order poles $z_1=i\pi/2$, $z_2=i\pi/2+2b\xi$, and in the limit $R\rightarrow\infty$ the integrals along the vertical paths $z=\pm R +i\phi$, $\phi \in [0,i\pi]$  vanish. From the residue theorem we have
\begin{equation}
\label{I:eq}
\oint_C f(z) dz=-i\pi I=2\pi i \sum_{k=1,2} Res f(z_k),
\end{equation}
where the residues can be calculated as usual:
\begin{eqnarray}
Res f(z_1)&=&\lim_{z\rightarrow z_1}\frac{d}{dz}(z-z_1)^2 f(z)=\frac{i\pi \cosh(2b\xi)}{\sinh^3(2b\xi)}+\frac{1}{\sinh^2(2b\xi)}\\
Res f(z_2)&=&\lim_{z\rightarrow z_2}\frac{d}{dz}(z-z_2)^2 f(z)=-\frac{(i\pi+4b\xi)\cosh(2b\xi)}{\sinh^3(2b\xi)}+\frac{1}{\sinh^2(2b\xi)}.
\end{eqnarray}
Combining this with \ref{I:eq} we have:
\begin{equation}
I=\frac{8\xi\cosh(2b\xi)}{\sinh^3(2b\xi)}-\frac{4}{b\sinh^2(2b\xi)}.
\end{equation}
\subsection{$I=\int_{-\infty}^{+\infty}\frac{\cos(2\mu x+\delta)}{\cosh(b(x+\xi))\cosh(b(x-\xi))}dx$}
Rewriting this with the definition $\omega=b(x+\xi)$ we have:
\begin{equation}
I=\frac{1}{b}\int_{-\infty}^{+\infty}\frac{\cos(\frac{2\mu\omega}{b})\cos(\delta-2\mu\xi)-\sin(\frac{2\mu\omega}{b})\sin(\delta-2\mu\xi)}{\cosh(\omega)\cosh(\omega-2b\xi)}d\omega,
\end{equation}
which can be expressed as
\begin{align}
I &=\frac{\cos(\delta-2\mu\xi)}{b}Re\left[ \int_{-\infty}^{+\infty}\frac{e^{i\frac{2\mu \omega}{b}}}{\cosh(\omega)\cosh(\omega-2b\xi)}d\omega \right] \nonumber \\
&\qquad {}-\frac{\sin(\delta-2\mu\xi)}{b}Im\left[ \int_{-\infty}^{+\infty}\frac{e^{i\frac{2\mu \omega}{b}}}{\cosh(\omega)\cosh(\omega-2b\xi)}d\omega \right].
\end{align}
We consider the following complex function integrated around $C$:
\begin{equation}
\oint_C f(z) dz=\oint_C \frac{e^{i\frac{2\mu z}{b}}}{\cosh(z)\cosh(z-2b\xi)}dz.
\end{equation}
Using the residue theorem we have:
\begin{equation}
\oint_C f(z) dz=\frac{(1-e^{-\frac{2\mu\pi}{b}}) }{b}\int_{-\infty}^{+\infty}\frac{e^{i\frac{2\mu \omega}{b}}}{\cosh(\omega)\cosh(\omega-2b\xi)}d\omega=2\pi i \sum_{k=1,2} Res f(z_k),
\end{equation}
and we can calculate the residues as before to find:
\begin{equation}
I=\frac{2 \pi \cos(\delta) \sin(2\mu\xi)}{b \sinh(\frac{\pi\mu}{b}) \sinh(2b\xi)}.
\end{equation}

\end{document}